\documentclass[11pt]{article}
\usepackage{epsfig,amsmath,amssymb}
\begin{document}

\title{\bf A Re-evaluation of Evidence for \\ Light Neutral Bosons \\ in Nuclear Emulsions \\}

\author{{F. W. N. de~Boer\footnote{Deceased}} \\
{\it LNBC, Amsterdam, The Netherlands} \\
\\
{C. A. Fields\footnote{Corresponding author: e-mail:fieldsres@gmail.com}}\\
{\it 21 Rue des Lavandi\`eres, Caunes Minervois, 11160 France}}
\maketitle

\begin{abstract}
Electron-positron pair-production data obtained by bombardment of emulsion detectors with either cosmic rays or projectiles with masses between one and 207 amu and kinetic energies between 18 GeV and 32 TeV have been re-analysed using a consistent and conservative model of the background from electromagnetic pair conversion.  The combined data yield a spectrum of putative neutral bosons decaying to $e^{+}e^{-}$ pairs, with masses between 3 and 20 MeV/$c^{2}$ and femtosecond lifetimes.  The statistical significance against background for these ``$X$-bosons'' varies between 2 and 8 $\sigma$.  The cross-section for direct production of $X$-bosons increases slowly with projectile energy, remaining over 1,000 times smaller than the pion production cross-section.  
\end{abstract}

\section{Introduction}

For over 50 years, measurements of $e^{+}e^{-}$ pair production by energetic 
ions incident on emulsion detectors have yielded events with opening angles 
at the $e^{+} - e^{-}$ vertex larger than those expected \cite{bors} for 
external pair conversion (EPC) by photons \cite{anand, hinter, jain74prot, 
jain74muon, nadi88, badawy89, kamel89, nadi96, kamel96, jain07}.  On the basis of 
relatively small data sets, El-Nadi and Badawy \cite{nadi88} and de Boer
 and van Dantzig \cite{boer88} proposed in 1988 that such events might
 represent the decays of neutral bosons with masses considerably larger
 than generally expected for Weinberg-Wilczek axions. Since then, 
additional studies of relativistic heavy-ion interactions with 
emulsion \cite{nadi96, kamel96, jain07, nagdy07} have been interpreted 
in terms of massive neutral boson production.  While such observations 
have triggered interest \cite{maddox88, perkins89, boer89}  
and are mentioned in the Review of Particle Physics' section on 
axion searches \cite{pdg08},  
no comprehensive attempt has 
been made to assess whether these observations represent new phenomena 
worthy of further investigation or background effects resulting from
well-characterized processes. 

In this paper, we revisit results from six accelerator studies using emulsion detectors, 
all of which report observations of $e^{+}e^{-}$ pairs that are interpreted as evidence for the existence of neutral bosons with masses between 1.5 and 
30 MeV/$c^{2}$ and lifetimes between $10^{-16}$ and $10^{-14}$ s \cite{jain74prot, jain74muon, nadi88, badawy89, kamel89,
nadi96, kamel96, jain07, nagdy07}.  Four of these studies \cite{nadi96, kamel96, jain07, nagdy07} were published after the 1988 proposal that such events might represent the decays of massive neutral bosons \cite{nadi88, boer88}, and hence can be considered tests of this proposal.  The assumptions made regarding background processes are not, however, consistent across these six studies, and the reported neutral boson masses are not identical; indeed Jain and Singh \cite{jain07} report a continuous spectrum of novel neutral bosons with masses up to 85 MeV/$c^{2}$.  We have therefore re-analysed these $e^{+}e^{-}$ pair production data using a single set of assumptions regarding background processes that are consistent with the experimental data reported by each of these studies.  We also re-analysed the classic cosmic-ray data of Anand \cite{anand} and Hintermann \cite{hinter} as comparisons.  The resulting 
background-subtracted data sets, taken together, raise the possibility
 of a spectrum of neutral bosons with energies between 3 and 20 MeV/$c^{2}$ 
and lifetimes on the order of femtoseconds. This mass-lifetime window,
 and in particular the existence of a 9.8 MeV/$c^{2}$ neutral boson decaying to
 $e^{+}e^{-}$ pairs, is consistent with measurements of anomalous
 internal pair production (IPC) in decays of excited states of 
light nuclei that suggest the existence of both unnatural parity (pseudoscalar and 
axial-vector) \cite{boer96, boer01, stiebing, kraszna05, kraszna06, vitez08} and 
natural parity (scalar and vector) neutral bosons \cite{fokke}. 

The existence of neutral particles decaying to $e^{+}e^{-}$ pairs in this 
mass-lifetime window was not investigated in early beam-dump measurements
\cite{davier, brown, riordan, bjorken}. 
Due to the long beam dumps, these experiments  
using high-energy electron and proton beams from laboratories including SLAC, Orsay and FNAL 
could only rule out lighter, longer-lived axion candidates 
(${\tau} \geq 10^{-9} s$). 
Results from a later experiment by Bross et al \cite {bross}, using a short beam dump   
with sensitivity to short-lived bosons with ${\tau} \leq 10^{-14} s$, 
partly overlap the present lifetime window but do not exclude it.  

If multiple neutral particles with similar rest masses exist, it is possible that their individual effects on the
 electron and muon anomalous magnetic moments would cancel out, potentially explaining the lack of evidence for massive axion-like particles in the available magnetic-moment data.  We conclude that the
 observations reviewed and re-evaluated here support the existence of fewer massive neutral particles than have been previously claimed \cite{nadi88, kamel89,
nadi96, kamel96, jain07, nagdy07}.  However, we interpret these data as providing sufficient suggestive evidence for novel massive neutral bosons with femtosecond lifetimes to encourage
 further experimental investigation.  If confirmed, the existence of such massive neutral bosons would provide \textit{prima facie} support for theoretical models that propose axion-like particles with 
masses in the MeV to 10s of MeV range \cite{fayet81, fayet07}, as are 
increasingly motivated by cosmological data \cite{boehm04, chen09}.

\section{Data and Analysis}

Fig.\,1 shows a compilation of $e^{+}e^{-}$ pair-production events from both 
cosmic-ray \cite{anand, hinter} and accelerator \cite{nadi88, kamel89, nadi96, badawy89, kamel96, jain07} data sets for which electron energy and opening angle data are 
available.  The signature of a massive particle $X$ decaying to $e^{+}e^{-}$ is an opening 
angle $\omega$ from the pair vertex larger than the Borsellino \cite{bors} angle $\omega_{P} =  4m_{e}c^{2}/E_{tot}$, where $m_{e}$ = 0.511 MeV/$c^{2}$ and $E_{tot}$ is the total pair 
energy in MeV, expected for EPC.  Emulsion detectors 
are ideal for detecting such events, as they allow accurate measurements of both the opening 
angle and the kinetic energies of the outgoing $e^{+}e^{-}$ pair.  

\begin{figure}[bt]
\centerline{\epsfig{file=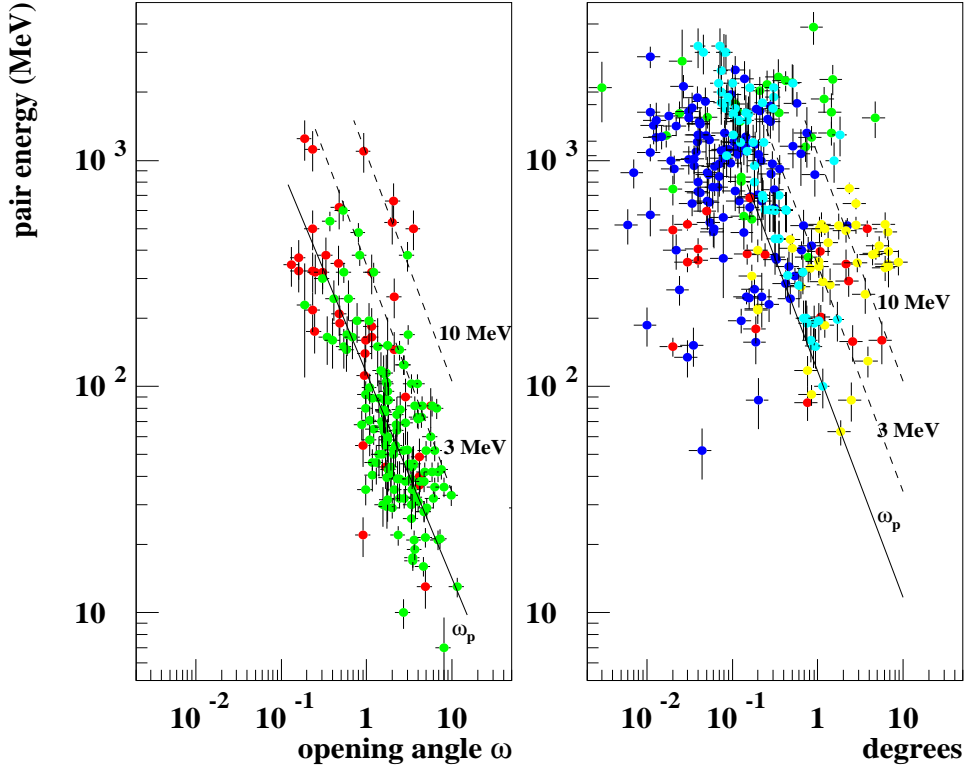,angle=0,width=5.0in,height=4.0in}}
\caption{
Total pair energy $E_{tot}$ versus opening angle $\omega$ in degrees for $e^{+}e^{-}$ pair-production 
events from both (a) low-energy (green \cite{hinter}) and high-energy (red \cite{anand}) 
cosmic-ray  and (b) 3.6 A${\cdot}$GeV (yellow  \cite{nadi88, kamel89}), 
60 A${\cdot}$GeV (red, 
\cite{badawy89}), 200 A${\cdot}$GeV indirect (blue \cite{nadi96}) and direct 
(green \cite{kamel96}), 
and 160 A${\cdot}$GeV (light blue \cite{jain07}) accelerator data sets, 
compared with 
Borsellino's \cite{bors} most probable opening angle $\omega_{P}$. 
For the 160 A${\cdot}$GeV $^{207}$Pb data set, only the 62 events reported to be above 
the Borsellino line in Fig.\,1f of \cite{jain07} are shown; the 1,158 events below the Borsellino line reported in this study have a distribution of total energies and opening angles very similar to thse reported for the 200 A${\cdot}$GeV $^{32}$S data set \cite{nadi96}.  Dashed lines labeled ``3 MeV'' and ``10 Mev'' indicate $e^{+}e^{-}$ opening angles corresponding \cite{bors} to the decays of 3 and 10 MeV/$c^{2}$ particles, respectively.
}
\label{fig:rawdata}
\end{figure}

As can be seen in Fig.\,1, a large fraction of the observed events fall below the Borsellino line.  All such events are consistent with $e^{+}e^{-}$ decay of a massless particle, i.e. with EPC by photons.  In the highest-energy reaction considered here, 160 A${\cdot}$GeV $^{207}$Pb on emulsion \cite{jain07}, 1,158 out of 1,220 recorded $e^{+}e^{-}$ events (95\%) have opening angles below the 
Borsellino line (cf. Fig. 1f of \cite{jain07}) and hence are \textit{prima facie} consistent with EPC; these points are not plotted in Fig. 1 as they would obscure the much smaller numbers of low-angle events recorded from lower-energy experiments.  It has been understood since the pioneering observations of Hintermann \cite{hinter} that the effective-photon approximation reflected by the Borsellino line significantly overpredicts the average opening angle and hence transverse momentum of EPC-derived $e^{+}e^{-}$ pairs at the pair-production threshold.  QED calculations of pair production cross-sections near threshold correctly reproduce both the observed transverse momenta and the near-threshold cross-section, confirming that the effective-photon approximation is accurate for apparent pair masses above the pair-production threshold \cite{adams04, klein04, bertulani}.

Previous analyses of these data have considered photons from $\pi^{0} \rightarrow 2 \gamma$ decay to be the primary source of the EPC background evident in Fig. 1 \cite{nadi88, kamel89, nadi96, jain07}; however, photons have a mean free path in emulsion of roughly 5 cm \cite{powell}, so photons from $\pi^{0}$ decay would be expected to efficiently escape the reaction zone. Indeed, Jain and Singh \cite{jain07} estimate that only 73 $e^{+}e^{-}$ pairs from $\pi^{0} \rightarrow 2 \gamma$ decay, distributed isotropically, would be observed within the area of emulsion that they analyzed following 160 A${\cdot}$GeV $^{207}$Pb bombardment, compared to 1,158 pairs with opening angles consistent with EPC actually observed within the forward cone only. A similar discrepancy between expected EPC from the $\pi^{0} \rightarrow 2 \gamma$ channel and observed events below the Borsellino line is evident in the analysis of the 200 A${\cdot}$GeV $^{32}$S reaction on emulsion \cite{nadi96, kamel96}. A reconsideration of background processes in these data sets is therefore necessary.

The production of photons as Bremsstrahlung radiation following Coulomb excitation of beam particles via interactions with target nuclei at large impact parameters has largely been neglected in previous analyses of these data sets, but can be expected to generate significant EPC backgrounds in these reactions.  The effective energy transfer of such interactions decreases by roughly an order of magnitude for every order of magnitude increase in impact parameter \cite{krauss}; the cross section for projectile excitations in the 100 keV to 10 MeV range, below the threshold for projectile breakup, is therefore expected to be on the order of 10,000 $\times$ the projectile breakup cross-section. Heavy nuclei Coulomb-excited below 10 MeV can be expected to de-excite by $\gamma$-emission, with average $\gamma$ multiplicities on the order of 10 per Coulomb excitation event.  Photons that are significantly aligned with the projectile momentum (${\sim}1/4$ of the total) will be Lorentz boosted to energies up to the maximum effective photon energy of $E_{max}(\gamma) \sim(0.03/A)E_{proj}$ \cite{bertulani}.  For 200 A${\cdot}$GeV $^{32}$S on emulsion, Baroni \textit{et al.} \cite{baroni90} report an average of 2.4 projectile breakup events per meter of penetration; hence one can expect about $2.4{\cdot}10^{5}$ $\gamma$'s due to Bremsstrahlung from Coulomb-excited projectiles per meter of target penetration for this reaction, with about $6{\cdot}10^{4}$ $\gamma$'s per meter in the forward cone.  The nuclear-interaction mean free path for 200 A${\cdot}$GeV $^{32}$S in emulsion is about 100 mm \cite{baroni90}, so one can expect on the order of 6,000 forward $\gamma$'s per observed nuclear interaction.  About half of these $\gamma$'s will convert within the first 5 cm from the breakup event; one would therefore expect about 3 EPC events per 50 $\mu$m downstream of an identified 200 A${\cdot}$GeV $^{32}$S - emulsion interaction.  In the case of the 160 A${\cdot}$GeV $^{208}$Pb data set, one would expect about 75 EPC events per 50 $\mu$m downstream of an identified interaction.  The experimental signature of such $\gamma$-interactions would be a nearly-uniform distribution of $e^{+}e^{-}$ events closely collimated to the pre-interaction projectile trajectory, attenuating linearly with distance $L$ from the projectile interaction as photons escape the viewing area of the tracking microscope.

Figure 2 shows the distribution of $e^{+}e^{-}$ events from 200 A${\cdot}$GeV $^{32}$S bombardment of emulsion as a function of distance $L$ from the observed projectile - emulsion reaction \cite{nadi96}.  Figure 2 can be compared with the very similar distribution of events from 160 A${\cdot}$GeV $^{207}$Pb on emulsion, as shown in Fig. 1(a) of \cite{jain07}.  In both of these studies, events very near the observed reaction were rejected; $L_{min} = 25 \mu$m for the 200 A${\cdot}$GeV $^{32}$S data \cite{nadi96} and $50 \mu$m for the 160 A${\cdot}$GeV $^{207}$Pb data \cite{jain07}.  Both $L$ distributions attenuate by roughly a factor of two per 1,000 $\mu$m over the entire range scanned, consistent with the 120 $\mu$m \cite{kamel89} and 100 $\mu$m \cite{jain07} fields of view of the scanning microscopes used and a Lorenz boost of roughly 1,000 $\times$ for forward-cone Bremsstrahlung photons.  These observed distributions support the interpretation of all events at $\geq 400 \mu$m from the reaction center in both data sets, most of which are below the Borsellino line, as due to Bremsstrahlung by Coulomb-excited projectiles.

\begin{figure}[bt]
\centerline{\epsfig{file=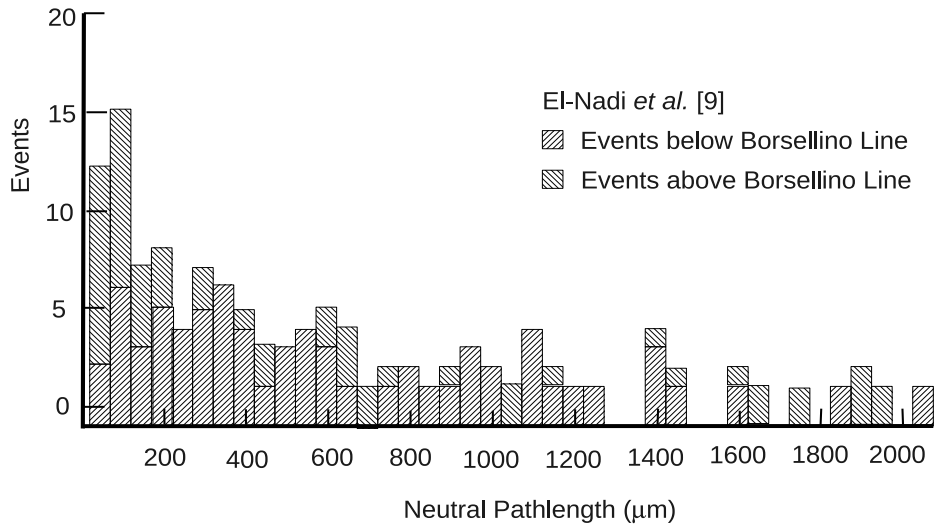,angle=0,width=5.5in,height=4.0in}}
\caption{
Distribution of $e^{+}e^{-}$ events from 200 A${\cdot}$GeV $^{32}$S on emulsion as a function of distance $L$ in $\mu$m from an identified breakup reaction center, as reported in \cite{nadi96}.  All points above the Borsellino line with $L \geq 600 \mu$m are from ``Cluster D'' in Table 1 and Fig. 6 of \cite{nadi96}.  The data are consistent with a uniform distribution of Bremsstrahlung-induced EPC events that attenuates as Bremsstrahlung $\gamma$ rays escape the viewing area of the tracking microscope.  
}
\label{fig:32Slength}
\end{figure}

Photons produced by Bremsstrahlung from Ag and Br nuclei in the emulsion target will not in general be Lorenz boosted, and hence will contribute negligibly to the number of EPC events observed in the vicinity of the beam track.  Reaction products including breakup fragments of the projectile (mainly $\alpha$ particles \cite{nadi96}) can be expected to be Coulomb-excited as they exit the emulsion; such fragments will produce Bremsstrahlung photons by the mechanism discussed above in proportion to their $Z^{2}$.  Approximately one photon per 2,500 $\mu$m of beam-fragment track from an identified projectile-target reaction is expected from this secondary mechanism in the case of 200 A${\cdot}$GeV $^{32}$S on emulsion.  However, heavy fragments typically exit the reaction zone at relatively high angles from the beam (cf. Figs 3-5 to 3-14 of \cite{kamel89} and Fig. 3 of \cite{nagdy07}), so Bremsstrahlung photons from beam-fragment excitation are not expected to contribute significantly to EPC events observed in the immediate vicinity the primary beam track.  Photons from $\pi^{0} \rightarrow 2 \gamma$ may contribute to EPC near the observed reaction center; however, the number of events observed at small $L$ compared to the large-$L$ tail in the 200 A${\cdot}$GeV $^{32}$S and 160 A${\cdot}$GeV $^{207}$Pb reactions does not appear to scale with the 2 $\times$ greater $\pi^{0}$ production expected \cite{hallman, dabrowska, adamovich} in the latter reaction.  Hence Bremsstrahlung from Coulomb-excited projectiles is expected to be the primary generator of EPC events in the data sets considered here.

The cross section for EPC by photons is strongly peaked at the production threshold value of $m(e^{+} + e^{-}) = 2m_e$, and decreases above threshold as $x/(1+x^{2})^{2}$, $x = \omega / \omega_{P}$, which for $E_{\gamma} \gg 2m_{e}c^{2}$ can be approximated as $1/m(e^{+} + e^{-})^{3}$ \cite{bors, klein04}.  For the 200 A${\cdot}$GeV $^{32}$S data, $E_{\gamma}{\sim}1$ GeV (Fig.\,5 
of \cite{nadi96}), while for the 160 A${\cdot}$GeV $^{207}$Pb data, $E_{\gamma} {\sim}2$ GeV (Fig.\,1f of \cite{jain07}); hence the above-threshold EPC background spectrum for these data sets can be expected to be identical up to normalization.  The 3.5 A${\cdot}$GeV $^{4}$He, $^{12}$C and $^{20}$Ne reactions generate photons with $E_{\gamma}{\sim}400$ MeV \cite{nadi88, kamel89}; the $1/m(e^{+} + e^{-})^{3}$ approximation is expected to somewhat over-predict the EPC background in this case, but is employed for consistency.  Fig.\,3 shows EPC background predictions for $m \geq 1.3$ MeV/$c^{2}$ for these three reactions, together with invariant mass projections for $m \geq 1.3$ MeV/$c^{2}$ for the 200 A${\cdot}$GeV $^{32}$S data from Table 1 of \cite{nadi96}, for the 160 A${\cdot}$GeV $^{207}$Pb data from Fig.\,1f of \cite{jain07} computed as described \cite{boer09}, and for the 3.5 A${\cdot}$GeV $^{4}$He, $^{12}$C and $^{20}$Ne data from Fig. 3-4 and Table 3-1 of \cite{kamel89}.  The lower cutoff of $m \geq 1.3$ MeV/$c^{2}$ is that chosen by El-Nadi \textit{et al.} \cite{nadi96} to reject threshold $e^{+}e^{-}$ production events, and is used to avoid under-prediction of EPC by the effective-photon approximation.  A bin size of 1 MeV/$c^2$ was employed consistently for all data; this bin size corresponds to the experimental mass resolution for pair masses between 3 and 7 Mev/$c^2$ reported by the studies for which tabulated data are available \cite{nadi88, badawy89, kamel89, nadi96, kamel96, nagdy07}.  

All of the data sets analyzed here represent selections from larger collections of observed by not analyzed events, hence the normalization of the EPC predictions, and in particular the ratio of threshold to above-threshold $e^{+}e^{-}$ production, cannot be calculated reliably from the available data.  Normalizations for the EPC predictions have therefore been chosen for each reaction to fit the data between 1.3 and 3.3 MeV/$c^{2}$, i.e. the first two bins of the histograms shown in Figs 3 and 4.  This choice of normalization is consistent with the well-established accuracy of the effective photon approximation above the pair-production threshold.  The alternative of normalizing all data sets at higher mass values would produce order-of-magnitude overpredictions of the low-mass data for which no plausible explanation could be offered.   The normalization of the EPC prediction for the 200 A${\cdot}$GeV $^{32}$S data with $L \geq 25 \mu$m shown in Fig. 3 is consistent with that used by Kamel \cite{kamel96} in analysing direct pair production ($L \leq 3 \mu$m) from 200 A${\cdot}$GeV $^{32}$S on emulsion.  The rejection of non-EPC events as background using this stringent normalization cannot be ruled out, especially for the lower-energy data sets.

\begin{figure}[hbt]
\centerline{\epsfig{file=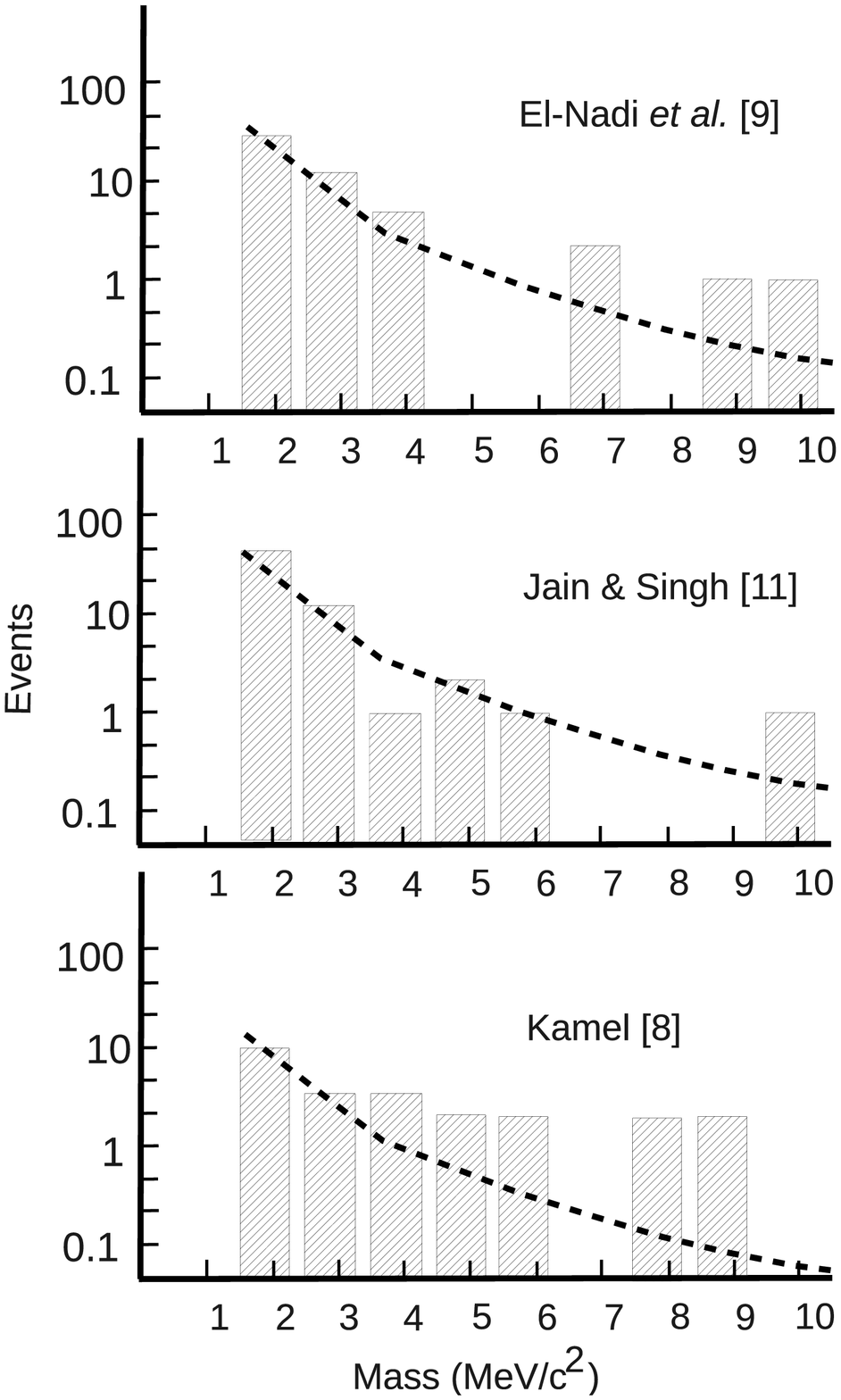,angle=0,width=4.5in,height=5.0in}}
\caption{
Background due to external pair conversion (EPC) induced by Bremsstrahlung in (top panel) 200 A${\cdot}$GeV $^{32}$S \cite{nadi96}, (middle panel) 160 A${\cdot}$GeV $^{207}$Pb \cite{jain07} and (lower panel) 3.5 A${\cdot}$GeV $^{4}$He, $^{12}$C and $^{20}$Ne \cite{nadi88, kamel89} reactions on emulsion.  Invariant masses for 200 A${\cdot}$GeV $^{32}$S are from Table 1 of \cite{nadi96}; invariant masses for 160 A${\cdot}$GeV $^{207}$Pb are computed from the total energy versus opening angle data presented in Fig. 1f of \cite{jain07} as described in \cite{boer09}; invariant masses for 3.5 A${\cdot}$GeV $^{4}$He, $^{12}$C and $^{20}$Ne are from Fig. 3-4 and Table 3-1 of \cite{kamel89}.  Masses are plotted in 1 MeV/$c^{2}$ bins starting at 1.3 MeV/$c^{2}$.  Curves in all panels are approximate effective photon (i.e. Borsellino) distributions calculated as $N/m^{3}$ where $m$ is the apparent rest mass of the $e^{+}e^{-}$ pair and $N$ is an normalization chosen to fit the data.
}
\label{fig:EPCback}
\end{figure}

As can be seen in Fig. 3, all of the events with invariant mass below 8 
MeV/$c^{2}$ reported in \cite{jain07} can be interpreted as due to EPC by 
Bremsstrahlung photons, as can events below 6 MeV/$c^{2}$ reported by \cite{nadi96} and events below 4 MeV/$c^{2}$ reported by \cite{nadi88, kamel89}.  The statistical significance $S/\sqrt{B}$, where $S$ represents counts in the ``signal'' and $B$ counts in the ``background'' \cite{hicks} varies from $\sigma = 2.2$ for the 9.8 MeV/$c^{2}$ event reported in \cite{jain07} to $\sigma = 6.3$ for the 8.8 MeV/$c^{2}$ event reported in \cite{nadi88, kamel89}.  Figure 4 shows background due to EPC for the cosmic-ray data sets of Hintermann \cite{hinter} and Anand \cite{anand} and the 60 A${\cdot}$GeV $^{16}$O data set of Badawy \cite{badawy89}, calculated and normalized using the same procedures as used in Fig. 3.  Similar to the data shown in Fg. 3 and those reported in \cite{kamel96}, these data sets reveal events above expected EPC background for $m \geq 3$ MeV/$c^{2}$.

\begin{figure}[hbt]
\centerline{\epsfig{file=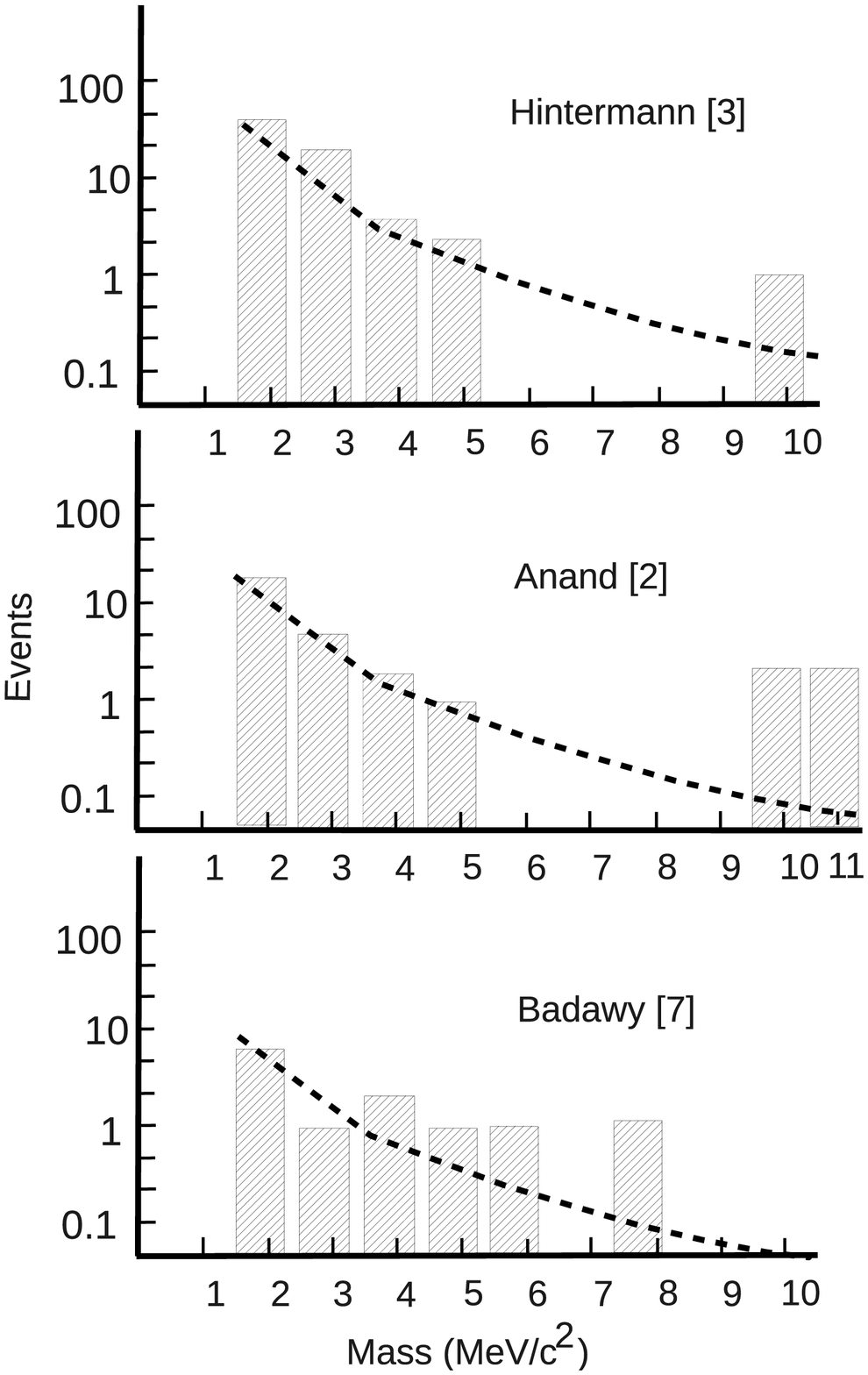,angle=0,width=4.5in,height=5.0in}}
\caption{
Background due to external pair conversion (EPC) induced by Bremsstrahlung in (top panel) cosmic ray data \cite{hinter}, (middle panel) cosmic ray data \cite{anand} and (lower panel) 60 A${\cdot}$GeV $^{16}$O reactions on emulsion \cite{badawy89}.  Masses are plotted as in Fig. 3.  Curves in all panels are approximate effective photon (i.e. Borsellino) distributions calculated as $N/m^{3}$ where $m$ is the apparent rest mass of the $e^{+}e^{-}$ pair and $N$ is an normalization chosen to fit the data.
}
\label{fig:EPCback2}
\end{figure}

Pair production by the Dalitz channel $\pi_{0} \rightarrow \gamma + (e^{+}e^{-})$ \cite{dalitz} has previously been deemed negligible in these data sets based on analyses of the energy partition asymmetries $y = (E_{2} - E_{1})/(E_{2} + E_{1})$, where $E_{2}$ is the greater of the two electron energies, across whole data sets \cite{nadi96, kamel96, jain07, boer88}.  The calculated EPC backgrounds shown in Figs 3 and 4, and in Fig. 2f of \cite{kamel96}, are therefore taken to represent the backgrounds due to known pair-production processes in these data sets.  Background subtraction was carried out for all spectra with a criterion of $\sigma = S/\sqrt{B} \geq 2.0$; the results of this subtraction are shown in Fig. 5.  These background-subtracted spectra are taken to be free of background events from characterized sources.

\begin{figure}[hbt]
\centerline{\epsfig{file=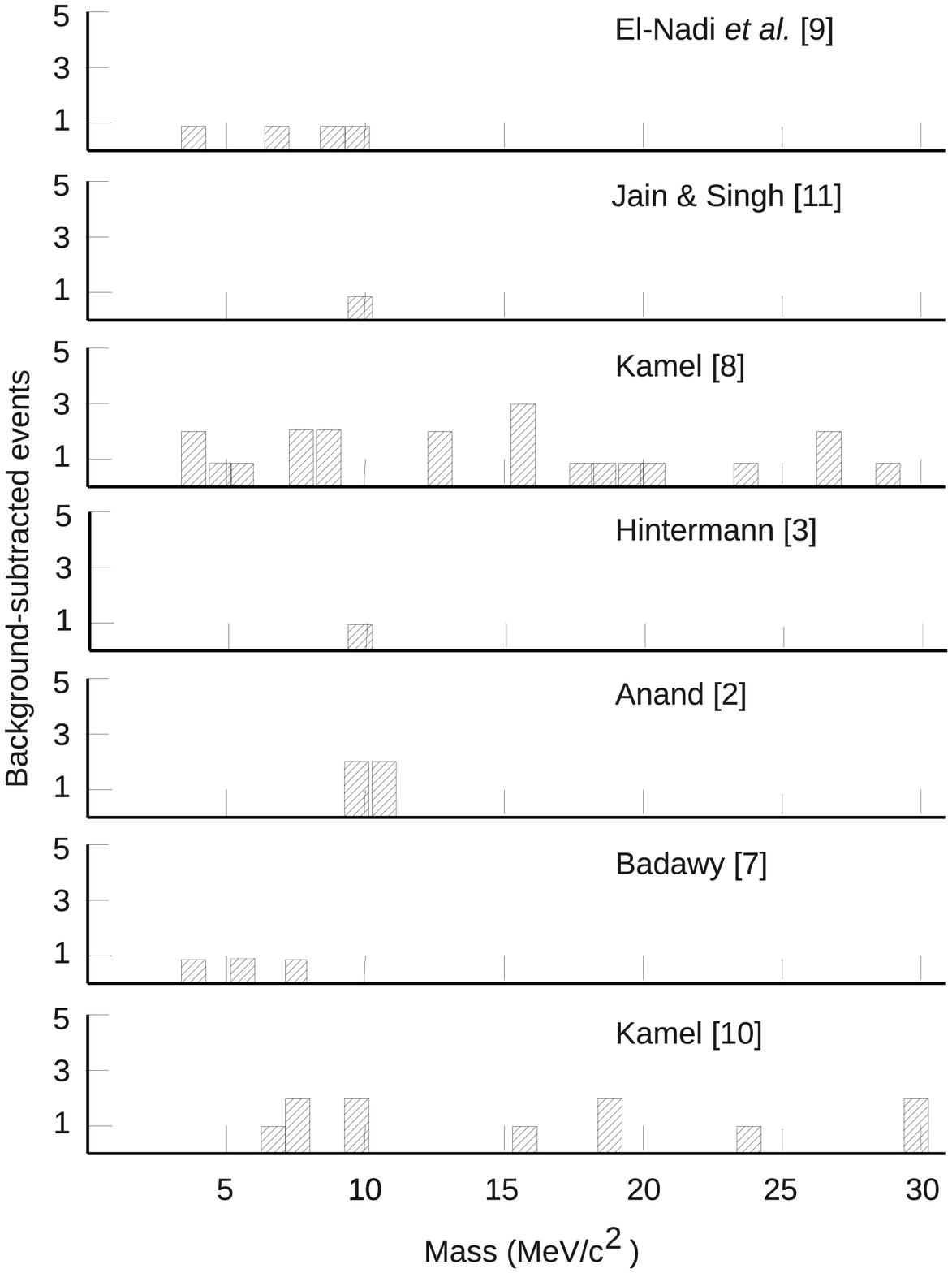,angle=0,width=6.0in,height=7.0in}}
\caption{
Background-subtracted data sets generated from Figs 3 and 4 above and Fig. 2f of \cite{kamel96} with a criterion of $\sigma = S/\sqrt{B} \geq 2.0$ where $S$ represents the observed events and $B$ the predicted EPC events for a given 1 MeV/$c^{2}$ mass bin.
}
\label{fig:back-sub}
\end{figure}

The events shown in the background-subtracted spectra of Fig. 5, together with one event at 9.8 MeV/$c^{2}$ well-separated from background from \cite{nagdy07}, events at 5.5, 7.0 and 16.5 MeV/$c^{2}$ well-separated from background from \cite{jain74prot}, and an event at 9.5 MeV/$c^{2}$ well-separated from background from \cite{jain74muon} were added within 1-MeV/$c^{2}$ bins; the sum spectrum is shown in Fig. 6.  While none of the studies from which these data are taken claimed to employ blind analysis methods \cite{klein-roodman05} and unknown effects due to measurement conditions or to inconsistent assumptions or analysis methods cannot be ruled out, the addition of these data sets to generate a single sum spectrum is justified if the published data are taken at face value.

The most straightforward interpretation of this sum spectrum is that it contains mass peaks with full width at half maximum (FWHM) of approximately 2 MeV/$c^{2}$ overlaid on a uniform background of approximately one event per bin.  The source of such a uniform background is unknown; however, there is no experimental motivation for assuming a nonuniform background, and in particular none for assuming a nonmonotonic background.  Assuming a FWHM of 2 MeV/$c^{2}$ and imposing a criterion of $\sigma = S/\sqrt{B} \geq 2.0$ against a uniform 1-event background, peaks can be identified at 3.8 MeV/$c^{2}$ (3$\sigma$), 5.8 MeV/$c^{2}$ (2.5$\sigma$), 7.8 MeV/$c^{2}$ (4.2$\sigma$), 9.8 MeV/$c^{2}$ (8.2$\sigma$), 15.8 MeV/$c^{2}$ (3$\sigma$) and 18.8 MeV/$c^{2}$ (2$\sigma$).  The 9.8 $\pm$ 1 MeV/$c^{2}$ peak is by far the most significant, and is supported by 8 of the 10 cited studies (all but \cite{badawy89} and \cite{jain74prot}).

\begin{figure}[hbt]
\centerline{\epsfig{file=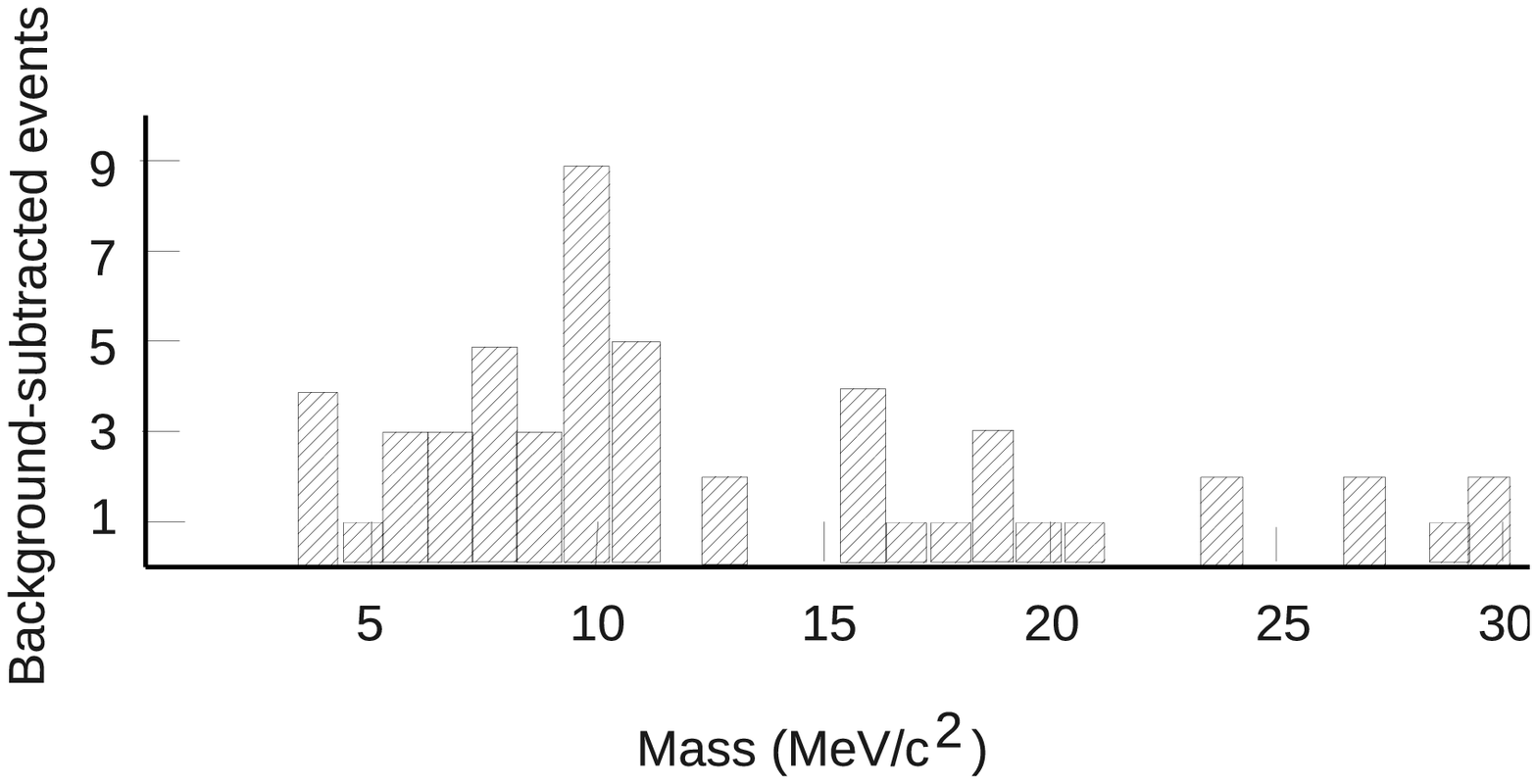,angle=0,width=6.0in,height=7.0in}}
\caption{
Sum of background-subtracted data sets from Fig. 5 together with five events well separated from background from other emulsion studies \cite{jain74prot, jain74muon, nagdy07}.
}
\label{fig:all-sum}
\end{figure}

The measurement conditions restrict the neutral particles producing these $e^{+}e^{-}$ events to the $10^{-16}$ and $10^{-14}$ s lifetime window.  Specific lifetime measurements are reported by \cite{nadi88, kamel89, nadi96, nagdy07}; reported lifetimes for the 3.8 $\pm$ 1 MeV/$c^{2}$ events are $4 \pm 1$ to $15 \pm 4 \times 10^{-16}$ s, for the 5.8 $\pm$ 1 MeV/$c^{2}$ events $3.6 \pm 0.9 \times 10^{-16}$ s, for the 7.8 $\pm$ 1 MeV/$c^{2}$ events $4.1 \pm 1.4$ to $6 \pm 2 \times 10^{-16}$ s, for the 9.8 $\pm$ 1 MeV/$c^{2}$ events $5 \pm 2$ to $31 \pm 9 \times 10^{-16}$ s, for the 15.8 $\pm$ 1 MeV/$c^{2}$ events $8 \pm 3$ to $13 \pm 2 \times 10^{-16}$ s, and for the 18.8 $\pm$ 1 MeV/$c^{2}$ events $9 \pm 5$ to $50 \pm 14 \times 10^{-16}$ s.  These measurements are consistent with decay halflives on the order of $10^{-15}$ s for the neutral particles producing these events.

\section{Discussion} 

The peaks identified in Fig. 6 can be interpreted as indicating the existence of a
 field $X$ outside the Standard Model, with excitations $X(m)$ that behave
 as neutral bosons with mass $m$.  The $X(m)$ have femtosecond lifetimes and
 decay to $e^{+}e^{-}$ pairs.  This interpretation of these events was first proposed in 1988 \cite{nadi88, boer88}; events consistent with X(9.8), in particular, were identified at that time by de Boer and van Dantzig \cite{boer88}.  Emission of a neutral boson consistent with X(9.8) 
has also been advanced as an explanation of anomalous IPC angular distributions in the 
decays of excited states of $^{4}$He, $^{8}$Be, and $^{12}$C \cite{boer96, boer01}, 
and well as the forbidden 10.96 MeV $0^{-} \rightarrow 0^{+}(g.s.)$ {\it M0} decay 
in $^{16}$O \cite{kraszna05,kraszna06}.  If the neutral bosons suggested by the 
 emulsion studies reviewed here may be identified with those suggested by IPC decay studies
 in light nuclei, $J^{P}$ assignments of $0^{-}$ and $1^{+}$ are
 indicated \cite{boer01, kraszna06, vitez08}.
The observation of putative $X$-bosons in nuclear decays
 \cite{boer01, kraszna06, vitez08} suggests that at least some of the
 $X$-bosons observed following cosmic ray or heavy-ion beam bombardment of
 emulsion targets may be produced by the decays of excited states of projectile fragmentation products.  Indeed, El-Nagdy et al. \cite{nagdy07} describe 8
 events from 200 A${\cdot}$GeV $^{32}$S on emulsion that are
 attributable on the basis of kinematic analysis to projectile-fragment
 decays.  If all secondary 200 A${\cdot}$GeV $^{32}$S events indicative of
 $X$ bosons are interpreted as projectile fragment decays, then assuming a
 maximum $Z = 2$ fragment multiplicity of 8, one obtains a minimal branching 
ratio $B_{X} = 3.7{\cdot}10^{-4}$ for the 1351 identified reactions observed 
\cite{nadi96}.  If all secondary 3.6 A${\cdot}$ GeV $^{12}$C and 
$^{22}$Ne events indicative of $X$ bosons are interpreted as projectile
 fragment decays, one obtains a minimal $B_{X} = 1.9{\cdot}10^{-4}$ for 
the 2600 identified reactions observed \cite{kamel89}. These values for 
$B_{X}$ are consistent with the range of $B_{X}$ values observed for 
individual states of light
 ${\alpha}$-nuclei, which assume a dominant $X \rightarrow e^{+}e^{-}$ decay
 branch for $X$ bosons \cite{boer01, kraszna06}.  
Secondary production of $X$ bosons by $\pi^{0}$ decay cannot be ruled out,
 but can be expected to be significantly less than the production of 
Dalitz channel $e^{+}e^{-}$ pairs, which are not observed in the present data sets \cite{boer88, nadi96, kamel96, jain07}. As $\pi^{0}$'s with kinetic 
energies greater than about 15 GeV would be expected to escape the detection
 area unless emitted at very forward angles, the contribution of a possible
 $\pi^{0} \rightarrow X$ channel is taken to be negligible.

The cross-section for direct production of $X$ bosons in nuclear interactions
 can be estimated by assuming that all background-subtracted $X$-boson events in which a particle decaying to an $e^{+}e^{-}$ pair appears to originate directly from the
 identified projectile interaction represent direct production. Total charged pion multiplicities $M(\pi^{\pm})$ \cite{nagdy07, hallman, dabrowska, adamovich} and $X$-boson multiplicities $M(X)$ calculated under this assumption
 are shown in Fig. 7.  The production of charged pions is well represented
 by a power-law fit, $M(\pi^{\pm}) = 1.3 K^{0.47}$ where $K$ is the kinetic
 energy of the projectile (cf. Fig.\,4 of \cite{nagdy07}).  The multiplicity of direct $X$ bosons \cite{kamel89, kamel96} is similarly well reproduced by a power law, $M(X) = 0.0022 K^{0.15}$.  Direct production of $X$ bosons thus appears to increase very slowly with beam energy, and is increasingly swamped by pion production as the available energy increases.

\begin{figure}[hbt]
\centerline{\epsfig{file=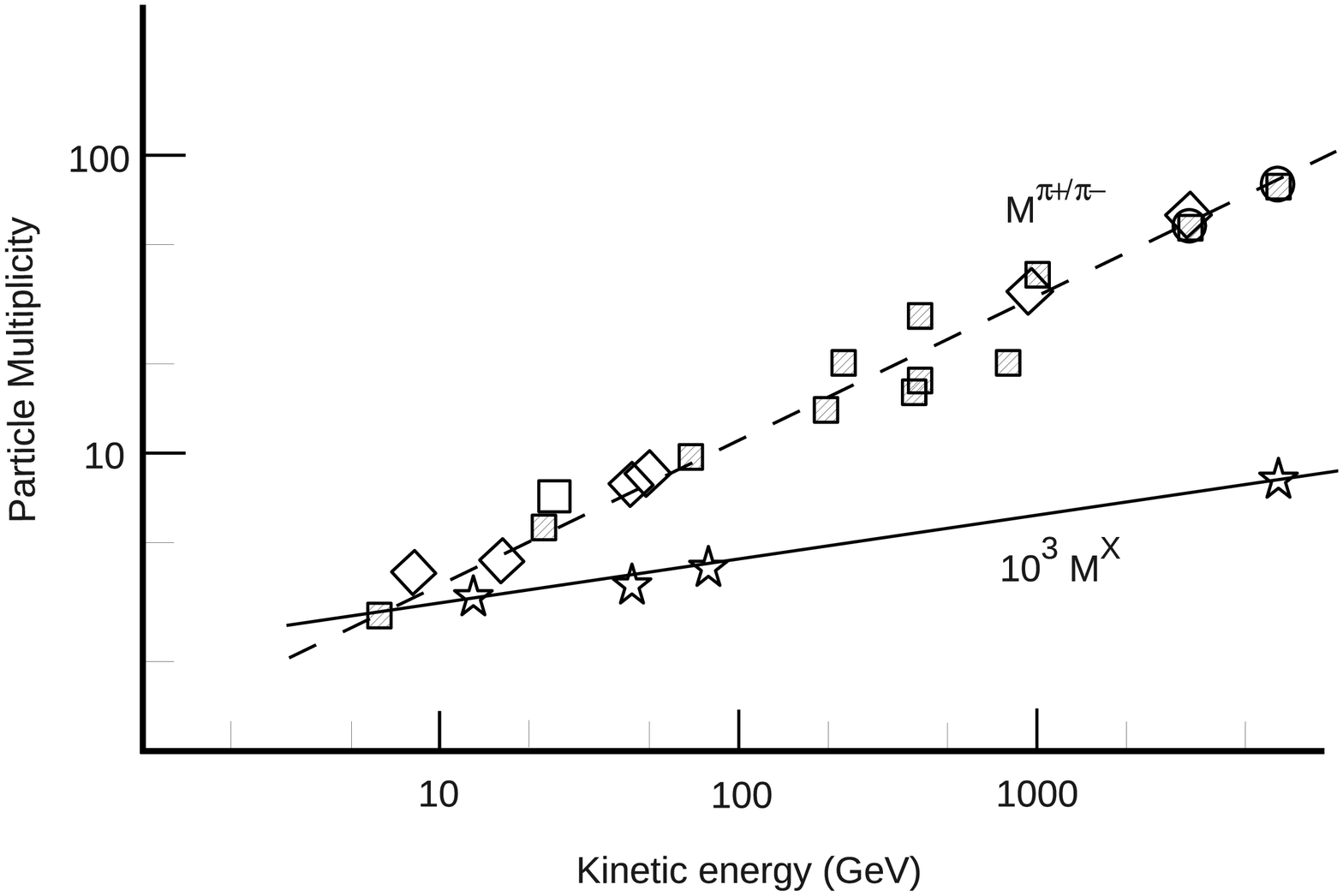,angle=0,width=6.0in}}
\caption{
Charged pion and $X$-boson multiplicities as functions of projectile kinetic
 energy $K$.  Pion data are from \cite{nagdy07} (open diamonds), \cite{hallman} (open square), \cite{dabrowska} (open circles) and \cite{adamovich} (hatched squares), and are fit with the power-law $M(\pi^{\pm}) = 1.3 K^{0.47}$ (dashed line).  $X$-boson data (open stars) are from \cite{kamel89, kamel96} and are fit with the power 
law $M(X) = 0.0022 K^{0.15}$ (solid line).  The plotted $X$-boson data have been multiplied by 1,000 for ease of visual comparison.
}
\label{fig:cross-sections}
\end{figure}

Any new particle decaying to $e^{+}e^{-}$ pairs can be expected to affect the anomalous 
magnetic moment  $(g-2)_{e}$ of the electron; assuming electron - muon universality, the anomalous magnetic moment $(g-2)_{\mu}$ would be affected as well \cite{reinhardt}. Values for the expected anomaly $\Delta a_{i} = \frac{1}{2} (g-2) - a_{QED}$, where $i$ ranges over scalar ($S$), pseudoscalar ($P$),
 vector ($V$) and axial-vector ($A$) couplings, calculated using the
 methods and form factors of Sch\"{a}fer \cite{schafer} are shown in
 Table 1,
 together with widths $\Gamma$ in eV.  As can be seen, any of the observed
 $X$-bosons considered individually would violate the existing constraints 
on the values of both $(g-2)_{e}$, i.e. $\Delta a_{i}= (1.24 {\pm} 0.95){\cdot}10^{-11}$, and $(g-2)_{\mu}$, i.e. $\Delta a_{i}=(27.5 {\pm} 8.4){\cdot}10^{-10}$ \cite{fayet07}. However, it is also clear that, within the relevant
 uncertainties on $X$-boson mass and width, anomalies contributed by bosons 
with different couplings could cancel out. Hence the current data imply 
that $X$-bosons must carry both parities for any given spin, i.e. that
 the pseudoscalar and axial-vector $X$-bosons implied by decay 
studies \cite{boer01, kraszna06, vitez08} must be accompanied by scalar 
or vector $X$-bosons of appropriate widths to cancel their contributions to
 the $\Delta a_{i}$.

\begin{table}
\scalebox{0.95}{
\begin{tabular}[htb]{|cccccc|}
\hline
$m_X$ & ${\Gamma}_{X}$  & ${\Delta} a^{e}_{S}$ & ${\Delta}a^{e}_{P}$ & ${\Delta}a^{e}_{V}$ & ${\Delta}a^{e}_{A}$ \\
\hline

$X(3.8)$ & $1.6 \pm 0.6$ & $8.3 {\cdot} 10^{-9}$ & $-7.8 {\cdot} 10^{-9}$ & $1.8 \cdot 10^{-9}$ & $-9.7 \cdot 10^{-9}$ \\
$X(5.8)$ & $1.8 \pm 0.5$ & $6.0 \cdot 10^{-9}$ & $-5.7 \cdot 10^{-9}$ & $1.3 \cdot 10^{-9}$ & $-7.0 \cdot 10^{-9}$ \\
$X(9.8)$ & $1.3 \pm 0.5$ & $7.6 \cdot 10^{-10}$ & $-7.5 \cdot 10^{-10}$ & $1.2 \cdot 10^{-10}$ & $-6.4 \cdot 10^{-10}$ \\
$X(15.8)$ & $0.8 \pm 0.4$ & $1.3 \cdot 10^{-10}$ & $-1.2 \cdot 10^{-10}$ & $1.8 \cdot 10^{-11}$ & $-9.0 \cdot 10^{-11}$ \\
$X(18.8)$ & $2.2 \pm 0.7$ & $1.9 \cdot 10^{-10}$ & $-1.9 \cdot 10^{-10}$ & $2.7 \cdot 10^{-11}$ & $-1.3 \cdot 10^{-10}$ \\
\hline
 &  & $\Delta a^{\mu}_{S}$ & $\Delta a^{\mu}_{P}$ & $\Delta a^{\mu}_{V}$ & $\Delta a^{\mu}_{A}$ \\
\hline
$X(3.8)$ & $1.6 \pm 0.6$ & $2.2 \cdot 10^{-7}$ & $-7.0 \cdot 10^{-8}$ & $1.4 \cdot 10^{-7}$ & $-5.6 \cdot 10^{-6}$ \\
$X(5.8)$ & $1.8 \pm 0.5$ & $1.6 \cdot 10^{-7}$ & $-5.1 \cdot 10^{-8}$ & $1.0 \cdot 10^{-7}$ & $-4.1 \cdot 10^{-6}$ \\
$X(9.8)$ & $1.3 \pm 0.5$ & $6.8 \cdot 10^{-8}$ & $-2.2 \cdot 10^{-8}$ & $4.5 \cdot 10^{-8}$ & $-1.3 \cdot 10^{-6}$ \\
$X(15.8)$ & $0.8 \pm 0.4$ & $2.5 \cdot 10^{-8}$ & $-8.4 \cdot 10^{-9}$ & $1.7 \cdot 10^{-8}$ & $-3.8 \cdot 10^{-7}$ \\
$X(18.8)$ & $2.2 \pm 0.7$ & $5.6 \cdot 10^{-8}$ & $-1.8 \cdot 10^{-8}$ & $3.6 \cdot 10^{-8}$ & $-7.8 \cdot 10^{-7}$ \\
\hline
\end{tabular}
}
\caption{ The predicted contributions of the observed $X$-bosons to the
anomalous magnetic moments of the electron $(g-2)_{e}$ and the muon $(g-2)_{\mu}$
assuming scalar $(S)$, pseudoscalar ($P$), vector ($V$) and axial-vector ($A$) character
for the bosons. The boson masses $m_X$ are listed in MeV/$c^{2}$ and the $X$-boson widths ${\Gamma}_{X}$ in eV.} 
\end{table}

The potential existence of $X$-bosons with the decay signature of ``heavy photons'' enables explanations of a number of anomalous results, in addition to those considered above, that have been observed in experiments measuring $e^{+}e^{-}$ pair production.  The JACEE Collaboration has already noted that three presumptive Bottom decays can be interpreted as producing $X$ bosons in the mass and lifetime range considered here \cite{jacee}.  In the early 1960's, Tsai-Ch\"u \textit{et al.} reported a presumptive \={n} annihilation event, now interpretable as an $\eta$ decay, producing three $\pi^{0}$ subsequently decaying to $e^{+}e^{-}e^{+}e^{-}$ \cite{tsai}.  The energies and divergence angles of the $e^{+}e^{-}$ pairs produced were sufficiently unusual that these events have been interpreted as evidence for a second neutral pion \cite{tsai, perkins01}.  As shown in Fig. 8, this event may instead be evidence for a $\pi^{0} \rightarrow 2X$ decay channel, producing one particle below the mass range observable above background here, $X(7.8)$, $X(9.8)$, and higher masses of 47, 66 and 102 MeV/$c^{2}$.  If this is the case, it would suggest the existence of $X$ bosons with $J^{P}$ of either $0^{+}$ or $1^{-}$ to conserve angular momentum and parity, consistent with the conclusion of the $(g-2)$ considerations outlined above.  While theoretical limits have been placed on the decay $\pi^{0} \rightarrow \gamma + X$ \cite{ignatiev}, a potential $\pi^{0} \rightarrow 2X, X \rightarrow e^{+}e^{-}$ channel was not considered, and the fraction of apparent $\pi^{0} \rightarrow e^{+}e^{-}e^{+}e^{-}$ decays that may proceed through $X$-boson intermediaries is unknown.  It is interesting in this regard that a significant peak at about 10 MeV/$c^{2}$ in the $e^{+}e^{-}$ spectrum from $\eta$ decay has been reported by the CLEO Collaboration (Fig. 2 of  \cite{lopez07}).  This peak was dismissed on the basis of GEANT simulations that indicate a significant background of single-$\gamma$ conversions following $\eta \rightarrow \gamma-\gamma$ decay, some of which escape the kinematic constraints used to identify true $\eta$ events \cite{lopez07, heltsley}.  However, the approximately 2 $\sigma$ difference at 10 MeV/$c^{2}$ between the data and GEANT simulation shown may indicate evidence for $X(9.8)$ in $\eta$ decay.  Anomalous events near this invariant mass have also been observed in 20 GeV 
photoproduction, but were dismissed as insufficiently statistically significant to warrant further study \cite{slac}.

\begin{figure}[hbt]
\centerline{\epsfig{file=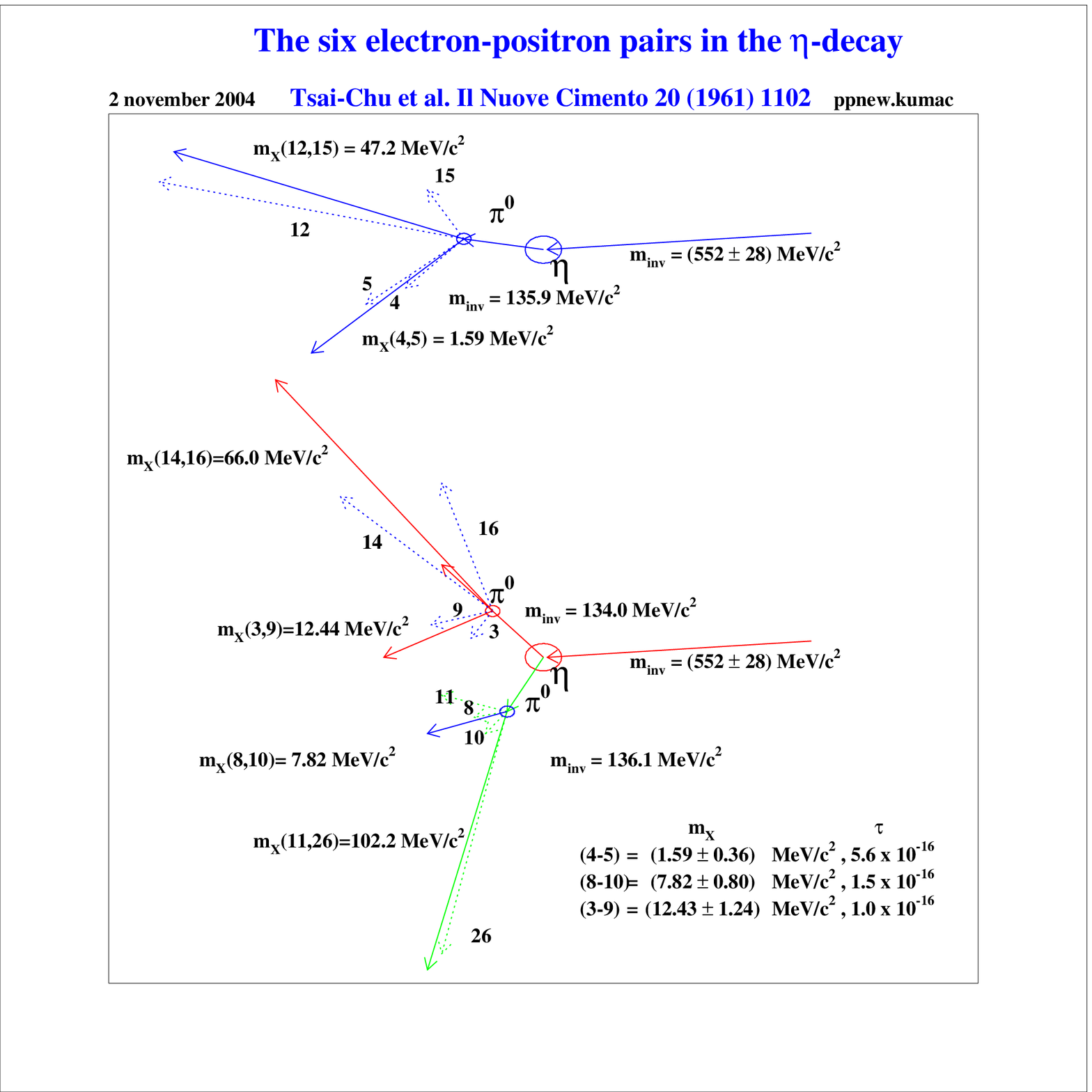,angle=0,width=4.6in}}
\caption{
Reconstruction of the kinematics of the presumptive $\eta$ decay reported by 
\cite{tsai} as involving $\pi^{0} \rightarrow 2X$ decays.
}
\label{fig:eta-decay}
\end{figure}

Two potential interpretations of the $X$-boson spectrum reported here can be
 suggested. One is motivated by Fayet's approach to $U1$-symmetry breaking
 through supersymmetry (SUSY) \cite{fayet81, fayet07}.  Fayet has proposed a
 spin-1 $U$-boson with double parity  and a mass and lifetime consistent with the $X$ bosons observed here, but with a dominant decay $U \rightarrow \nu \bar{\nu}$ and a small coupling to leptons.  The observed $X$ bosons do not 
satisfy this latter requirement. However, a condensate of $U$-bosons could, 
in principle, exhibit both a spectrum of masses with both natural and 
unnatural $J^{P}$ values, and a larger coupling to the $e^{+}e^{-}$ channel. 
Preliminary analyses of anomalies in the angular distributions of $e^{+}e^{-}$
 decays of excited states in the ${\alpha}$-nuclei $^{8}$Be, $^{12}$C 
and $^{16}$O that assume such a condensate of very light ($\sim114$ keV) 
bosons suggest that such a mechanism could produce the observed $J^{PC}$ and
 transition energies.

A second line of interpretation is suggested by the relation between $X$-boson and pion production discussed above (Fig.\,7). Assuming bare $u$ and $d$ quark masses in the MeV range \cite{pdg08}, one can envision the production
 of weakly-interacting $(u,\bar{u})$ or $(d,\bar{d})$ pairs, in the absence
 of gluons, with masses and $J^{P}$ in the range observed here. 
Such excitations would exist in the low-energy limit of QCD, for which the
 shapes and strengths of quark potentials are unknown. As indicated by Fig.\,7, production of such excitations would be strongly suppressed, relative to pions, above the threshold for gluon production, so all quarks would appear bound by gluons,
 i.e. as Standard Model mesons or hadrons, in high-energy experiments.

\section{Conclusions}

We have reviewed evidence from heavy-ion bombardment and cosmic-ray 
studies using emulsion detectors that suggest the existence of a 
spectrum of neutral $X$-bosons with masses between 3 and 20 MeV/$c^{2}$ and
 lifetimes on the order of femtoseconds.  With a conservative and consistent model of the photo-production background imposed on all studies reviewed, 5 putative $X$-bosons, with masses of 2.3, 5.8, 9.8, 15.8 and 18.8 MeV/$c^{2}$ are suggested by the combined data.  The statistical significance of
 individual $X$-boson peaks from the combined studies, relative to an assumed uniform one event per Mev/$c^2$ background of unknown source, varies between 2 and 8 $\sigma$. The derived branching ratios for secondary production of such $X$ bosons by projectile fragments is 
consistent with branching ratios for putative $X$-bosons observed in decay 
studies of light nuclei \cite{boer01, kraszna06}. Under our assumptions, the derived direct 
production multiplicity for such $X$-bosons increases very slowly with 
projectile kinetic energy as $M(X) = 0.0022 K^{0.15}$ across 
all projectile energies investigated, indicating that  
$X$-boson production is much smaller than pion production at all energies above the pion production threshold. Consideration of the predicted effects of such 
$X$-bosons on the anomalous magnetic moments of both the electron and muon 
rules out the possibility of a single massive $X$-boson, but leaves open the possibility that $X$-boson excitations include both natural and 
unnatural parity states for each spin.

We conclude that the experimental observations reviewed here offer evidence for the existence of
new neutral bosons with masses between 3 and 20 MeV/$c^{2}$ and femtosecond lifetimes that encourages dedicated further searches for 
such particles. The inferred behaviour of the direct production cross-section
indicates that such searches are best carried out below or in the vicinity of
the pion production threshold.

\section*{Author's Note}
Fokke W. N. de Boer (1942 - 2010) passed away as this work was nearing completion.  This paper is dedicated to his memory.

\section*{Acknowledgment}
The comments of an anonymous referee on a previous version of this paper significantly contributed to its clarity.

\end{document}